\documentclass[aps,preprint,final,prb,12pt]{revtex4}
\usepackage{times}
\usepackage{graphicx}
\begin{document}

\title{
Large magnetoresistance in bcc Co/MgO/Co and
FeCo/MgO/FeCo tunneling junctions
\footnote{The submitted manuscript has been authored by a contractor of the
U.S. Government under contract No. DE-AC05-00OR22725.  Accordingly, the
U.S. Government retains a nonexclusive, royalty-free license to publish
or reproduce the published form of this
contribution, or allow others to do so, for U.S. Government
purposes.}
}
\author{
X.-G. Zhang$^1$ and W. H. Butler$^{2,3}$}
\affiliation{1. Computer Science and Mathematics Division,
Oak Ridge National Laboratory, Oak Ridge, TN 37831\\
2. Metals and Ceramics Division,
Oak Ridge National Laboratory, Oak Ridge, TN 37831\\
3. Center for Materials for Information Technology,
University of Alabama, Tuscaloosa, AL 35487-0209
}

\date{\today}
\begin{abstract}
By use of first-principles electronic structure calculations, we predict
that the magnetoresistance of the bcc Co(100)/MgO(100)/bcc Co(100) and
FeCo(100)/MgO(100)/FeCo(100) tunneling
junctions can be several times larger than the very large magnetoresistance
predicted for the Fe(100)/MgO(100)/Fe(100) system.  The origin of this large
magnetoresistance can be understood using simple physical arguments
by considering the electrons at
the Fermi energy travelling perpendicular to the interfaces.
For the minority spins there is no state with $\Delta_1$ symmetry whereas
for the majority spins there is only a $\Delta_1$ state.  The $\Delta_1$ state
decays much more slowly than the other states within the MgO barrier.
In the absence of scattering which breaks the conservation of
momentum parallel to the interfaces, the electrons travelling
perpendicular to the interfaces undergo total reflection if the moments of the
electrodes are anti-parallel.  These arguments apply equally well to systems
with other well ordered tunnel barriers and for which the most slowly decaying
complex energy band in the barrier has $\Delta_1$ symmetry.  Examples include
systems with (100) layers constructed from
Fe, bcc Co, or bcc FeCo electrodes and Ge, GaAs, or ZnSe barriers.

\end{abstract}
\maketitle
\newpage


Spin dependent tunneling junctions formed by trilayers of FM/insulator/FM
 or FM/semiconductor/FM where FM represents a ferromagnet have been shown to
have relatively large magnetoresistance\cite{Parkin1,Parkin2,Moodera1,Moodera2}.
Much larger magnetoresistances have been predicted in recent calculations on
epitaxial spin tunneling systems\cite{LKKR,FeMgO,FeOMgO,Mathon,Dederichs}.
We have earlier\cite{FeMgO} calculated the electronic structure and
the spin-dependent tunneling conductance of Fe(100)/MgO(100)/Fe(100) sandwiches, using
the first-principles layer-KKR\cite{LKKR} approach. It was found that for
this system, the majority spin conductance when the two Fe layers
are aligned is dominated by contributions
from $k_\parallel=0$, {\it i.e.} by electrons travelling perpendicular
to the interface.   The importance of the complex energy
bands within the barrier in determining the rate of decay of electrode Bloch states
of similar symmetry\cite{FeMgO,Dederichs} was also emphasized.  This effect
of wave function symetry was explained in terms
of symmetry induced oscillations of the wave function in the plane of the
interface\cite{lat-var}.  In addition, these
calculations predicted a number of surprising phenomena such as quantum interference
between tunneling states.  Recently, calculations performed independently by Mathon and
Umerski\cite{Mathon} have also predicted very large TMR for the Fe(100)/MgO(100)/Fe(100)
system.

These earlier works predicted TMR ratios as high as 6000\%(defined as
the ratio of the change in resistance to the parallel resistance).
These are much larger than the ratios that have been been reported\cite{Vincent,Bowen,Yuasa}
to date which are approximately 100\% or less.
It is believed that at least part of the reason for the observed TMR being lower
than that calculated is a strong affinity between Fe and O which causes
a partial layer of FeO to form at the interface between Fe and MgO\cite{FeO,Meyerheim}.
Recently, we have performed calculations which showed that an FeO layer would
indeed, dramatically lower the TMR\cite{FeOMgO}.


In this paper we consider the symmetric junctions bcc Co(100)/MgO(100)/Co(100) and
FeCo(100)/MgO(100)/FeCo(100) with epitaxial lattices.
In the latter system, the electrodes are formed by an ordered FeCo bcc
alloy. Effects of disorder are not included in the present calculations.
It is possible that Co(100)/MgO(100) or FeCo(100)/MgO(100) may be easier to grow
without an interfacial transition metal oxide layer than Fe(100)/MgO(100).
Indeed, preliminary reports of large TMR in systems of the form FeCo(100)/MgO(100)/FeCo(100)
and FeCo(100)/MgO(100)/AlGaAs(100) have been presented\cite{reports,Parkin}. In these
cases, FeCo seems to have been a crystalline bcc alloy with at
least some substitutional disorder.

 Similarly
to our previous calculations\cite{FeMgO}, the electrode layers of bulk bcc
cobalt are fixed at the lattice constant of 2.82 \AA, while the lattice
constant of bulk bcc FeCo is chosen to be 2.86 \AA. The MgO lattice
constant is taken to be a factor of $\sqrt{2}$ larger than that of the bulk
electrodes,
therefore the (100) layers of the two materials can be matched epitaxially.
We assume no vertical relaxations between the layers.
In order to fill the space with a minimal amount of
overlap between the spheres, an empty sphere is inserted
between the MgO and the electrode layers at the MgO/Co or MgO/FeCo interface.

The self-consistent calculation is performed in the same manner as in
Refs.\cite{FeMgO, FeOMgO}. We limited our calculations within magnetic configuration space
in the sense that all electron spins are assumed collinear.  We also assumed that
the magnetic order has the same periodicity as the two-dimensional lattice thus
disallowing antiferromagnetic ordering within the same atomic layer. Antiferromagnetic
coupling between layers is allowed, however.  The tunneling conductance is calculated using the same approach as in
Ref.\cite{FeMgO}. It uses the LKKR code which
implements the Landauer-B\"uttiker conductance formalism\cite{Landauer,Buttiker}
within the first-principles KKR framework.

\begin{figure}[tbp]
\includegraphics[angle=0,width=0.6\textwidth]{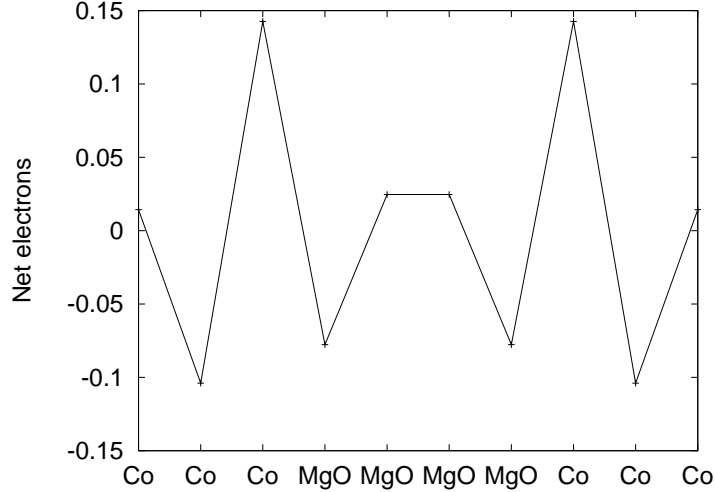}
\caption{Net charge on each atomic layer of the Co/MgO/Co tunnel junction.}
\label{charge}
\end{figure}

We find that the charge rearrangement necessary to correctly
offset the bands of the MgO relative to those of Co leads to very little
charge transfer between layers (Fig. \ref{charge}), similar to the result we
obtained for the Fe/MgO interface. We also found little charge transfer
at the FeCo/MgO interface.


The tunneling conductance for the three types of electrodes, bcc Fe(100), bcc(Co)100 and
B2 FeCo(100) all using an 8ml MgO(100) barrier are shown in Table \ref{cond_table}.
Although our physical arguments emphasize $k_\parallel=0$, it should be emphasized
that the results presented in Table \ref{cond_table}, resulted from an integral over the
entire two dimensional Brillouin zone with 8256 $k$-points in $1/8$th of the
zone.
For the minority channel, interfacial resonance states generate extremely sharp
peaks as a function of ${\bf k_\parallel}$.  The contributions from these peaks have been
omitted because they are difficult to calculate accurately.  If they had been included,
the calculated TMR would have been slightly higher because for the Co and FeCo electrodes
they contribute to the minority parallel conductance but do not contribute significantly
to the anti-parallel conductance.

\begin{table}
\caption{Tunneling conductivity (in $1/\Omega m^2$) for various spin channels for the Co/MgO/Co
and FeCo/MgO/FeCo tunnel junctions. Each junction contains 8 atomic layers of
MgO. Resonant state contributions to the
minority spin channel have been removed. Results for Fe/MgO/Fe
are also listed for comparison. The electrode materials are all assumed to have
the bcc phase and all interfaces are normal to the (100) direction.}
\begin{tabular}{ccccc}
\colrule
Spin alignment&up-up&down-down&up-down (down-up)&$\sigma_P/\sigma_{AP}$\\
\colrule
FeCo/MgO/FeCo&$1.19\times 10^9$&$2.55\times 10^6$&$1.74\times 10^6$&353.5\\
\colrule
Co/MgO/Co&$8.62\times 10^8$&$7.51\times 10^7$&$3.60\times 10^6$&147.2\\
\colrule
Fe/MgO/Fe&$2.55\times 10^9$&$7.08\times 10^7$&$2.41\times 10^7$&54.3\\
\end{tabular}
\label{cond_table}
\end{table}

\begin{figure}
\includegraphics[angle=0,width=0.6\textwidth]{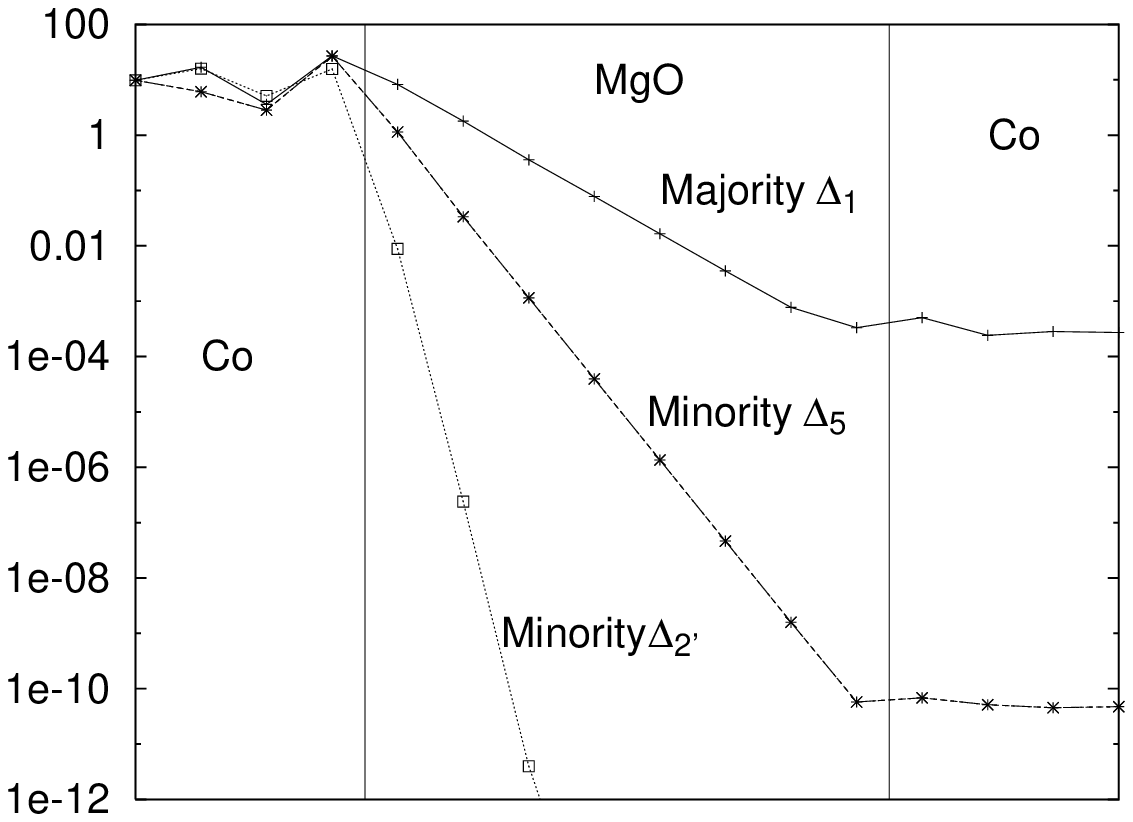}
\includegraphics[angle=0,width=0.6\textwidth]{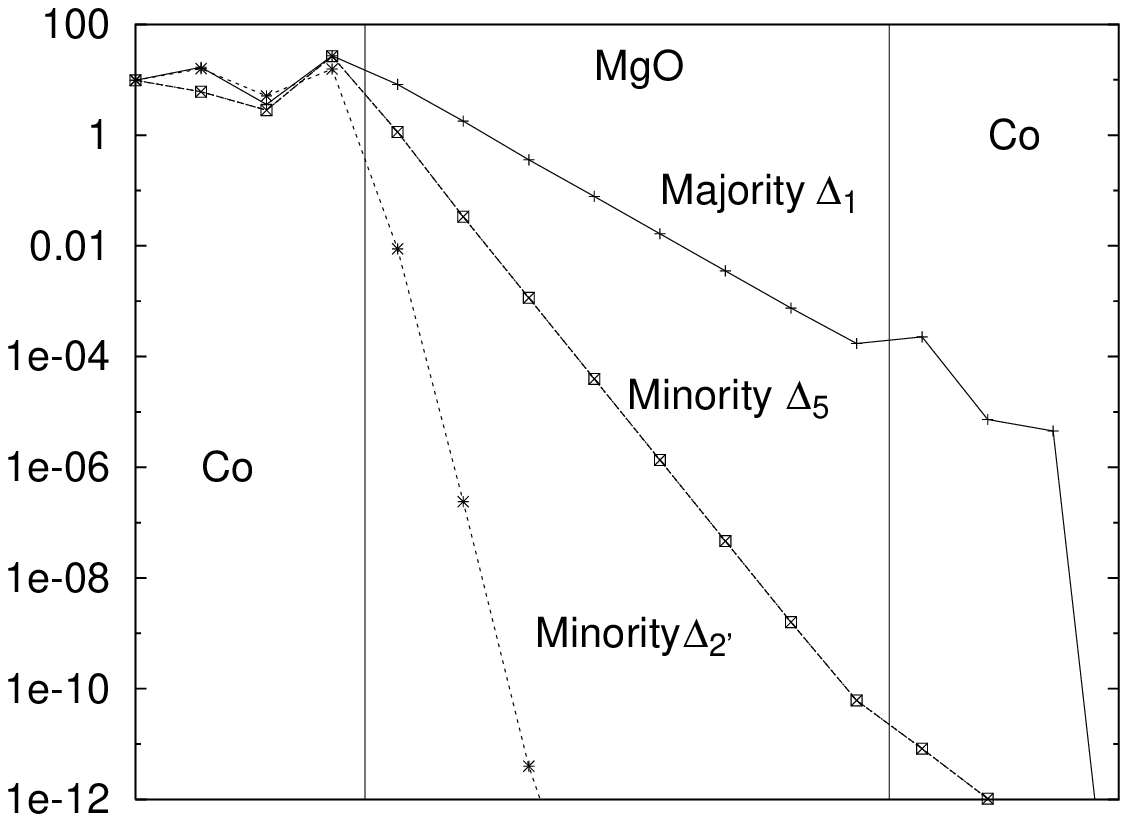}
\caption{Tunneling density of states on each atomic layer at $k_\parallel=0$ for
Co/MgO/Co tunnel junction. Top panel, parallel spin alignment,
bottom panel, antiparallel spin alignment}
\label{CoTDOS}
\end{figure}

\begin{figure}
\includegraphics[angle=0,width=0.6\textwidth]{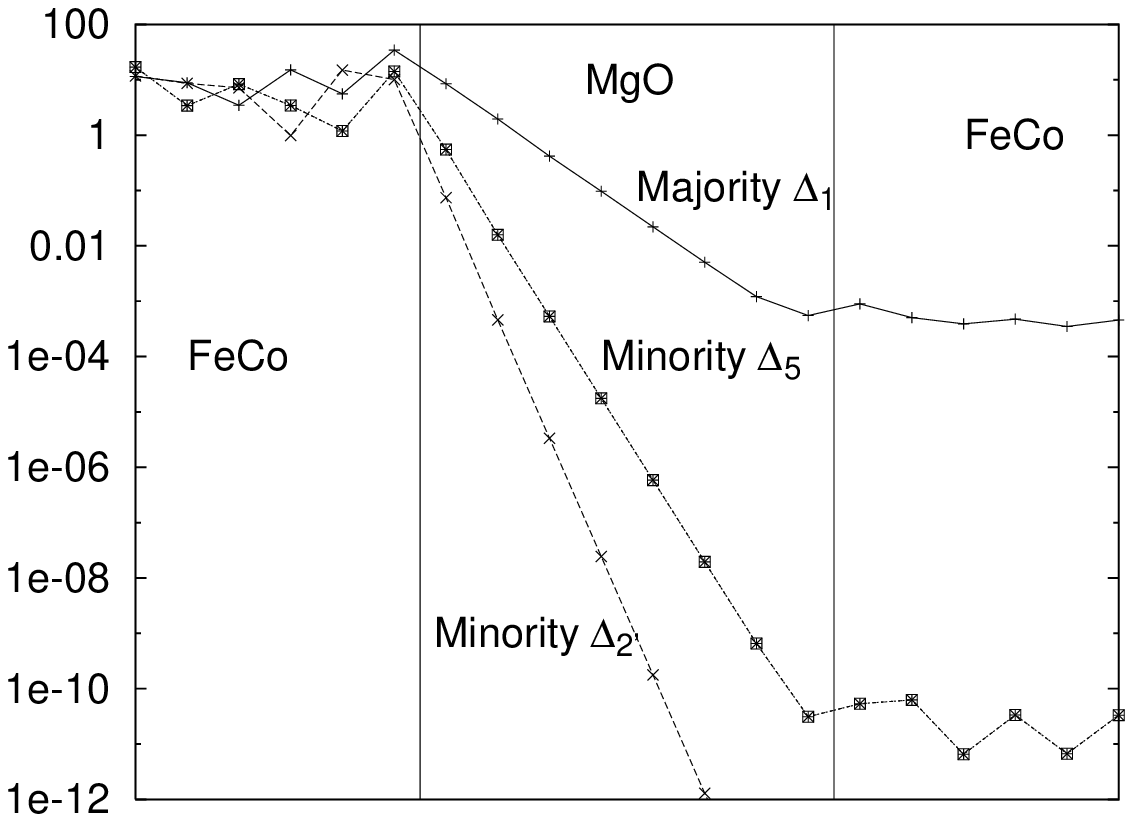}
\includegraphics[angle=0,width=0.6\textwidth]{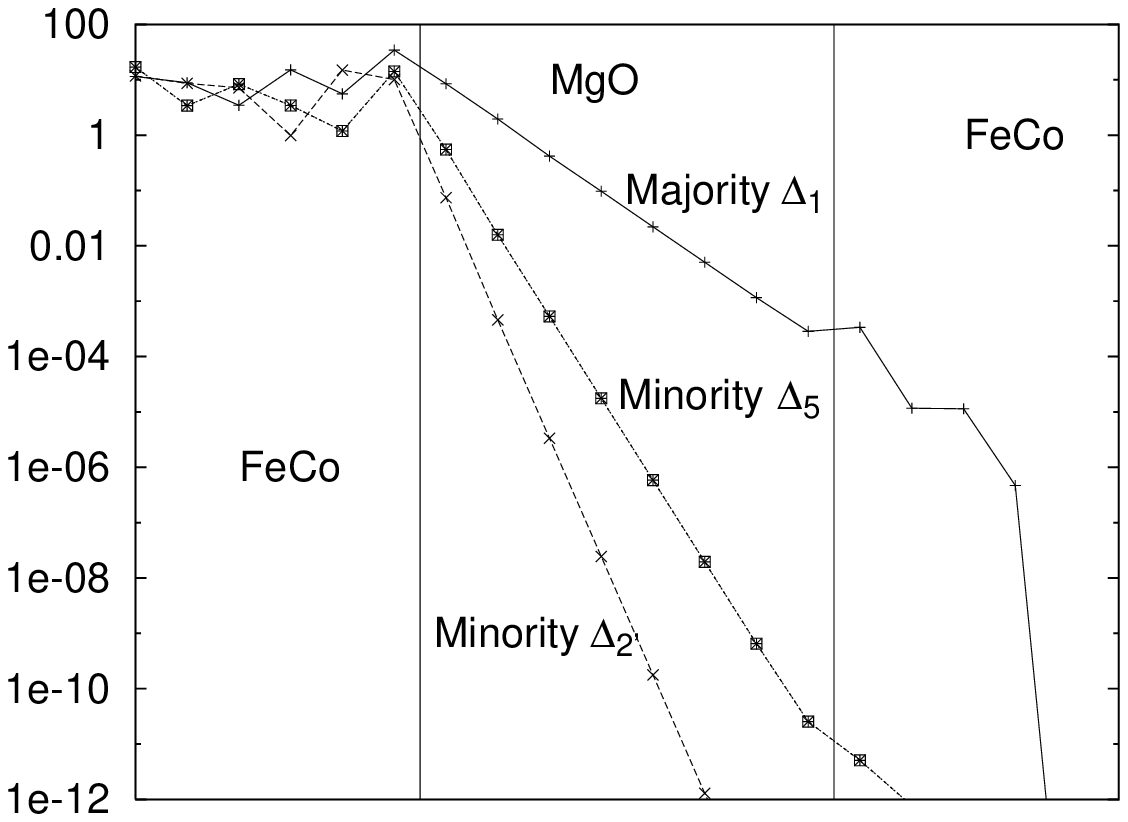}
\caption{Tunneling density of states on each atomic layer at $k_\parallel=0$ for
FeCo/MgO/FeCo tunnel junction. Top panel, parallel spin alignment,
bottom panel, antiparallel spin alignment}
\label{FeCoTDOS}
\end{figure}

The tunneling density of states (TDOS) is defined as
the electron density of states at
each layer due to a single incident Bloch state from the left Fe (Co or CoFe) lead.
On each layer, the TDOS
is roughly proportional to the modular square of the wave function that
matches to the incident Bloch state.
The TMR is dominated by the parallel
majority spin conductance, which in turn is dominated by the contribution
from $k_\parallel=0$. As shown in Figs. \ref{CoTDOS} and \ref{FeCoTDOS},
the majority spin TDOS for the
$\Delta_1$ state decreases much more slowly in the MgO layer than the
states with different symmetries.  This is true for both
the Co/MgO/Co stack and the FeCo/MgO/FeCo stack and is similar to the result
we obtained for the Fe/MgO/Fe tunnel
junction. A striking feature that distinguishes Co/MgO/Co and FeCo/MgO/FeCo
from earlier results for Fe/MgO/Fe, is that for the antiparallel
spin alignment, {\it all} states are completely reflected
at $k_\parallel=0$. This effect is the reason for the much larger conductance
ratio for Co/MgO/Co and FeCo/MgO/FeCo than Fe/MgO/Fe tunnel junctions.

\begin{figure}
\includegraphics[angle=0,width=0.6\textwidth]{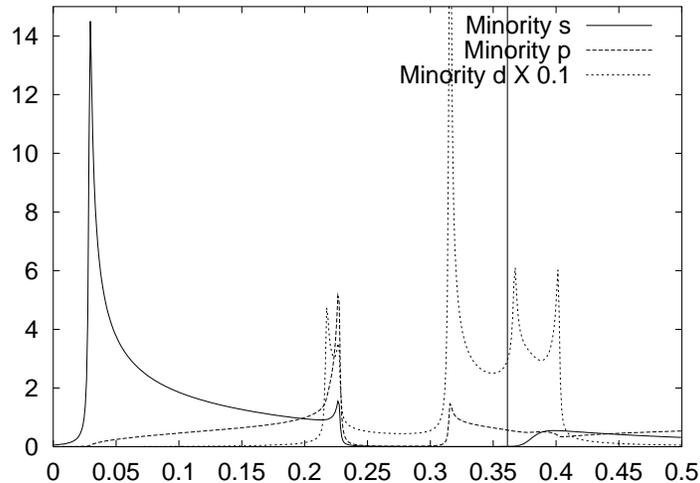}
\caption{Partial density of states at $k_\parallel=0$ in bulk bcc cobalt.}
\label{Cospd}
\end{figure}
\begin{figure}
\includegraphics[angle=0,width=0.6\textwidth]{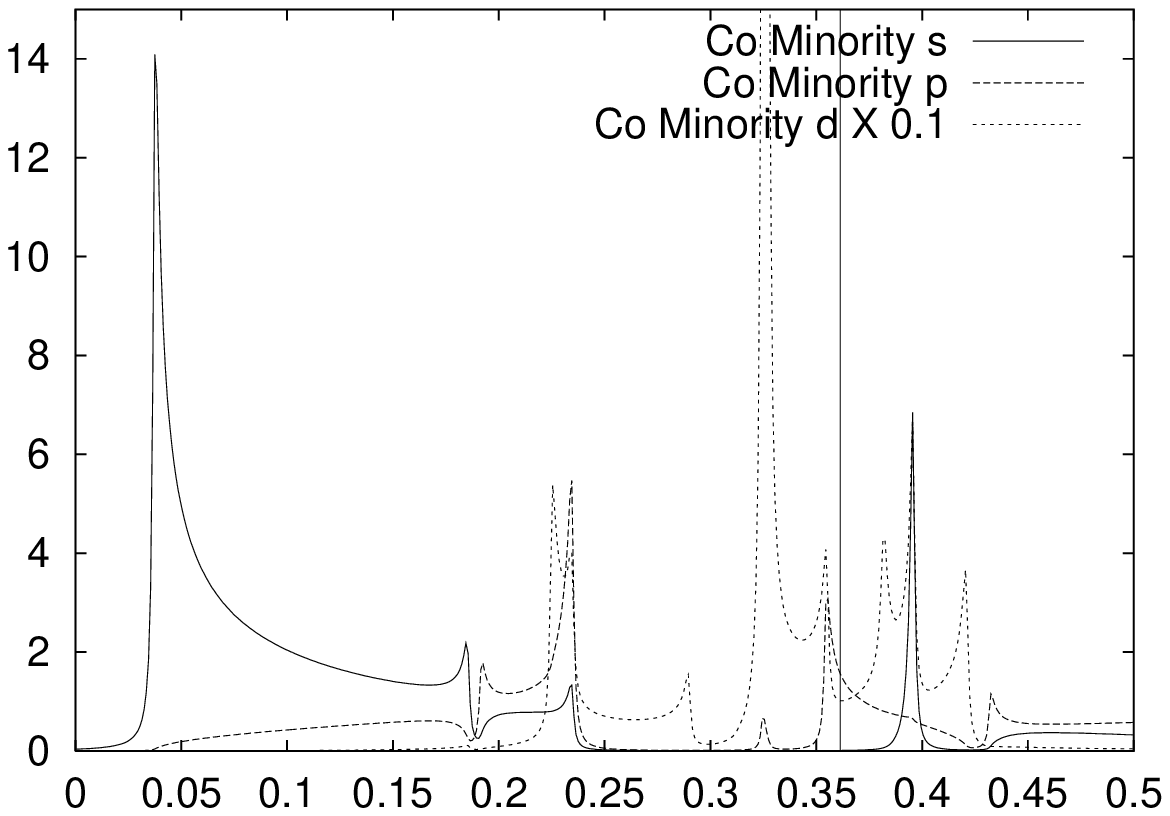}
\includegraphics[angle=0,width=0.6\textwidth]{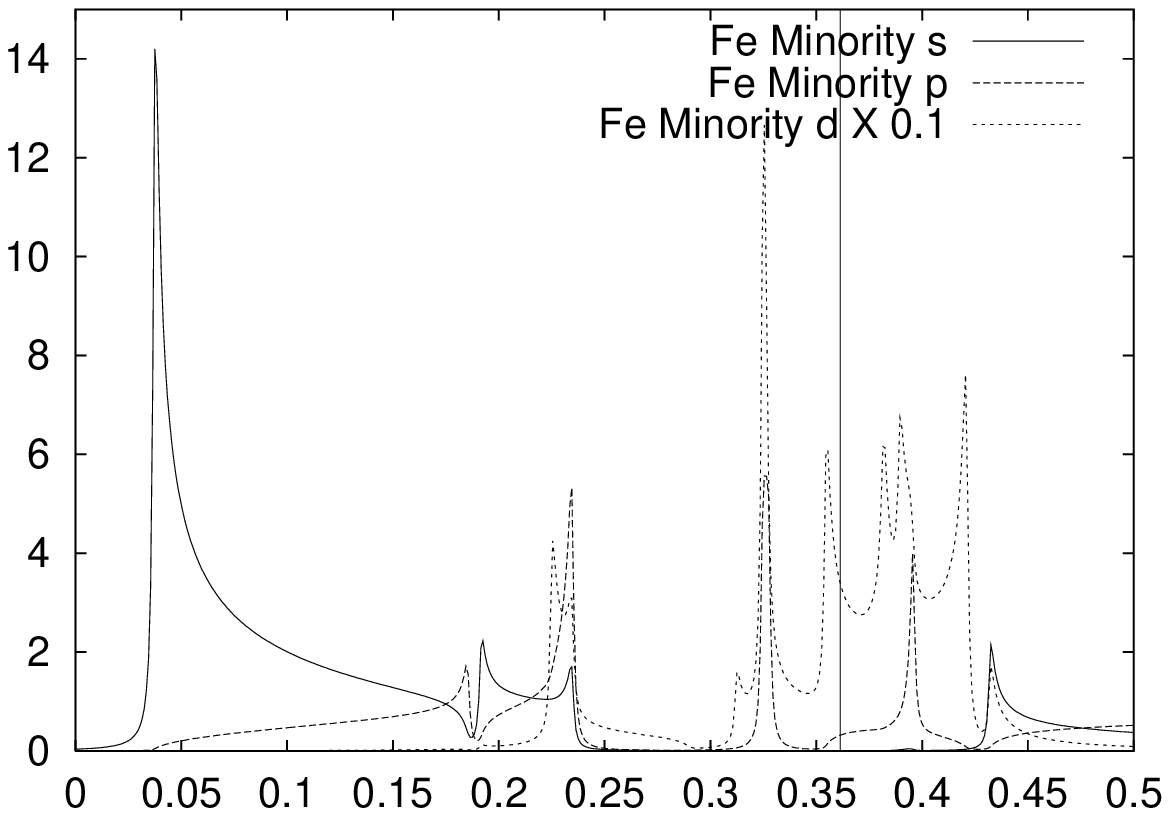}
\caption{Partial density of states at $k_\parallel=0$ in bulk bcc FeCo.
Top panel, Co sublattice, bottom panel, Fe sublattice}
\label{FeCospd}
\end{figure}
The total reflection of the tunneling electrons at $k_\parallel=0$
for antiparallel spin alignment is due to the fact that there are no Bloch eigenstates
in the minority spin channel with $\Delta_1$ symmetry.  This results from
the hybridization of the ``$s$-band'' with the $d$-bands.  In a generic
bcc transition metal electronic structure the $s$-band typically starts from
the $\Gamma$ point a few eV below the $d$-bands.  Its energy rises rapidly with $k$ until it
encounters the $d$-bands at which point it flattens out and ends at the Brillouin
zone boundary near the bottom of the $d$-band complex.  Starting above the $d$-band complex,
the ``$s$-band'' can also be followed downward in energy.  It is again highly dispersive
until it approaches the $d$-band complex at which point it flattens out and intersects
the zone center at the $\Gamma_{2'}$ point.  In the (100) direction, the ``$s$-band''
is the one with $\Delta_1$ symmetry and there is a range of energy over which there is
no $\Delta_1$ band.  For bcc Co and bcc FeCo, the spin splitting is such that
no $\Delta_1$ band crosses the minority electron Fermi energy and the only band
that crosses the majority Fermi energy is a $\Delta_1$ band.

A consequence
of this is that the minority states at $k_\parallel=0$ have no $s$ component.
Only the $\Delta_1$ wave functions are compatible with $\ell=0$ symmetry
when expanded about an atomic center.
The absence of $s$-DOS is evident in the plots of the partial DOS at $k_\parallel=0$
of the minority spin electrons in Figs. \ref{Cospd} and
\ref{FeCospd}. In both plots, the Fermi energy (indicated by the vertical bar)
falls within the energy gap of the $s$ partial DOS. Because all of the slowly
decaying tunneling states
within the MgO layer must have $s$ component due to the symmetry of the
wave function in MgO, a Bloch state with zero $s$ component decays very rapidly
in the MgO layer. Conversely, in the absence of any Bloch states with a
nonzero $s$ component, a tunneling state can not exit the MgO layer.

Half-metallic ferromagnetic electrodes, ({\it i.e.} ferromagnets with states
of only one spin channel at the Fermi energy) are not required in order to
obtain very large TMR.
If one
can achieve sufficiently good two dimensional periodicity within the barrier and
near the interface that $k_\parallel$ is reasonably well conserved, {\it i.e.} the
scattering is mostly specular, then one may take advantage of a class of electrode-barrier
combinations in which some of the states of one spin channel decay much more slowly in the
barrier than those of the other. In addition, a total reflection of the electrons
traveling perpendicular to the interface when the moments of the electrodes
are antiparallel can further increase the TMR by several fold.

This argument is, of course, more speculative in the presence of disorder.  Some types of
disorder, however, may not completely eliminate the effect.  For strong magnetic alloys, {\it i.e.}
those with filled majority $d$-bands, the moments are such that the bands match extremely well
in the majority channel.  Therefore, $k_\parallel$ conservation arguments can be applied to majority
electrons.  The problem will be in the minority channel where the scattering is
expected to be relatively strong.  Even there however, the bcc Co and FeCo electrodes
should offer the possibility for relatively large TMR.  Consider the case of anti-parallel
alignment.  Majority electrons injected from the left electrode (as in the lower pannels of
Figures \ref{CoTDOS} and
\ref{FeCoTDOS}) will decay slowly in the MgO barrier.  When they encounter the right
electrode, however, the (initially) $\Delta_1$ states that would decay exponentially if the
electrode were well ordered will, we speculate, continue do so for several layers until diffuse
scattering converts a significant fraction of the surviving flux into symmetries that can propagate.

Finaly, we note that the total reflection depends on the
absence of the $\Delta_1$ band for the minority spin in the cubic (100)
direction. The fact that the measurements of FeCo/MgO/FeCo junctions\cite{Parkin}
showed
greatly reduced TMR for the (110) textured samples and for non-bcc Co rich electrodes
confirms that the role of Co is more than simply preventing the formation of an
FeO layer, and the total reflection of the $\bar\Gamma$ electrons may play
a crucial role in achieving the high TMR.


This work was supported by the Office of Basic Energy
Sciences Division of Materials Sciences of the U.S. Department of Energy.
Oak Ridge National Laboratory is operated by UT-Battelle, LLC, for
the U.S. Department of Energy under contract DE-AC05-00OR22725.  Work
at the University of Alabama was supported in part by National
Science Foundation MRSEC Grant Number DMR0213985 and by the Office
of Naval Research.

\end{document}